\documentclass[journal=ancac3,manuscript=article]{achemso}

\usepackage{chemformula} 
\usepackage[T1]{fontenc} 
\usepackage{multirow}

\author{Qilin Guo}
\affiliation[SHU]
{Department of Physics, Shanghai University, Shangda Road 99, 200444 Shanghai, China}

\author{Yuriy Dedkov}
\affiliation[SHU]
{Department of Physics, Shanghai University, Shangda Road 99, 200444 Shanghai, China}
\email{dedkov@shu.edu.cn}

\author{Elena Voloshina}
\affiliation[SHU]
{Department of Physics, Shanghai University, Shangda Road 99, 200444 Shanghai, China}
\email{voloshina@shu.edu.cn}

\title[Intercalation of Mn in graphene/Cu(111) interface]
  {Intercalation of Mn in graphene/Cu(111) interface: Insights to the electronic and magnetic properties from theory}

\begin{document}




\begin{abstract}
The effect of Mn intercalation on the atomic, electronic and magnetic structure of the graphene/Cu(111) interface is studied using state-of-the-art density functional theory calculations. Different structural models of the graphene-Mn-Cu(111) interface are investigated. While a Mn monolayer placed between graphene and Cu(111) (an unfavorable configuration) yields massive rearrangement of the graphene-derived $\pi$ bands in the vicinity of the Fermi level, the possible formation of a Cu$_2$Mn alloy at the interface (a favorable configuration) preserves the linear dispersion for these bands. The deep analysis of the electronic states around the Dirac point for the graphene/Cu$_2$Mn/Cu(111) system allows to discriminate between contributions from three carbon sublattices of a graphene graphene layer in this system and to explain the bands' as well as spins' topology of the electronic states around the Fermi level. 
\end{abstract}

\section{Introduction}
\label{S:Intro}

Since the discovery of the unique properties of graphene~\cite{Novoselov:2004it}, many fascinating phenomena in this material were demonstrated, exemplified by massless Dirac fermion physics~\cite{Geim:2007hy,Katsnelson:2006kd,Huard:2007ia}, an anomalous quantum Hall effect~\cite{Novoselov:2005es,Zhang:2005gp,Novoselov:2006hu,McCann:2006ev}, and superconductivity~\cite{Cao:2018ff}. As an ideal 2D material, graphene has many possible attractive practical applications, such as protective coatings~\cite{Weatherup:2015cx}, touch screens~\cite{Bae:2010es,Ryu:2014fo}, gas sensors~\cite{Schedin:2007} and many others. With respect to the graphene-metal systems, the synthesis of graphene on metals is considered as one of the most promising technological approach~\cite{Bae:2010es}, with further processing of the obtained graphene layers. Moreover, the graphene-metal interfaces, particularly graphene-ferromagnet, are proposed for the realization of the spintronics applications based on the transport of electrical charge or/and spin, such as the spin filters~\cite{Karpan:2007tf} and stable spin emitters~\cite{Dedkov:2008d,Dlubak:2012fq}. Therefore, the understanding of the crystallographic structure and electronic properties of such graphene-metal interfaces~\cite{Dedkov:2015kp,Yang:2020fda} is an initial prerequisite for the further studies of these systems.

From the point of view of electronic structure, graphene adsorbed on metal surfaces shows two distinct cases~\cite{Voloshina:2012c}: (1) in the \textit{weakly} bonded case, graphene is always doped ($n$-type or $p$-type), while the electronic structure characteristic for freestanding graphene remains almost intact [examples: graphene/Ir(111)~\cite{Pletikosic:2009}, graphene/Pt(111)~\cite{Klimovskikh:2017}, graphene/Al(111)~\cite{Khomyakov:2009}, and graphene/Cu(111)~\cite{Gottardi:2015}] and (2) in the \textit{strongly} bonded case, graphene overlayer demonstrates the significant distortion of the linear dispersion relation for the graphene $\pi$ bands due to the strong overlap of the graphene states and valence band states of the underlying materials (space, energy, and $k$-vector overlapping), that leads to the complete disappearance of the Dirac point in graphene [examples: graphene/Ni(111)~\cite{Bertoni:2005}, graphene/Co(0001)~\cite{Eom:2009}, graphene/Ru(0001)~\cite{Marchini:2007}].

The electronic structure of graphene on metals can be altered in different ways, like intercalation of different species in graphene/metal interfaces, adsorption of atoms or molecules on top of graphene, graphene's edge engineering, creation of the graphene nanostructures, etc. It is well known that the intercalation between graphene and metallic substrates can strongly affects the properties of the graphene-metal interface. There are several possible results of this process: (1) The intercalated species may decouple graphene from strongly interacting substrates, such as in the case of graphene/Al/Ni(111)~\cite{Voloshina:2011jn,Generalov:2013gi}, graphene/Au/Ni(111)~\cite{Voloshina:2018ga} and graphene/Cu/Ni(111)~\cite{Dedkov:2001hg}; (2) Such intercalated layers lead to the change the carrier concentration in graphene, and even change the carrier type (from holes to electrons) such as in the case of Cu intercalation in graphene/Ir(111)~\cite{Vita:2014aa}; (3) Intercalated metals may also enhance the magnetic coupling between a ferromagnetic substrate and graphene, such as graphene/Fe/Ni(111)~\cite{Weser:2011hd} and graphene/Ni$_3$Mn/Ni(111)~\cite{Voloshina:2019ei}, which improves the possible spin-filtering properties of graphene; (4) The intercalated layer in itself may bring new properties to graphene, such as in the case of intercalated lithium, where superconductivity in graphene has been predicted to occur~\cite{Profeta:2012hg}.

Considering that graphene can be synthesized on Cu foil and the electronic properties of \textit{weakly}-bonded graphene on Cu(111) has been well studied, the intercalation of Mn atoms (which possess the high magnetic moment due to the half-filled $d$-shell, Mn $3d^5$) in the graphene/Cu(111) system could be an interesting way to tailor the electronic and magnetic properties of graphene. Here, we present systematic structural and electronic properties studies of the system formed after intercalation of Mn in the graphene/Cu(111) interfaces using the density functional theory (DFT). It is found that the intercalation of a monolayer of Mn and formation of the sharp graphene/Mn/Cu(111) interface changes the electronic properties of graphene greatly, while the formation of the Cu$_2$Mn interface alloy on Cu(111) only makes graphene $n$-doped and leaves the linear dispersion of the graphene-derived $\pi$ states at the Fermi level ($E_F$), which is similar to the case of graphene on Cu(111). The magnetic properties of the obtained interfaces and their possible application in graphene-based spin filtering devices are discussed in details.

\section{Results and discussion}

\label{S:Results}

\begin{figure}[h]
\centering\includegraphics[width=0.5\textwidth]{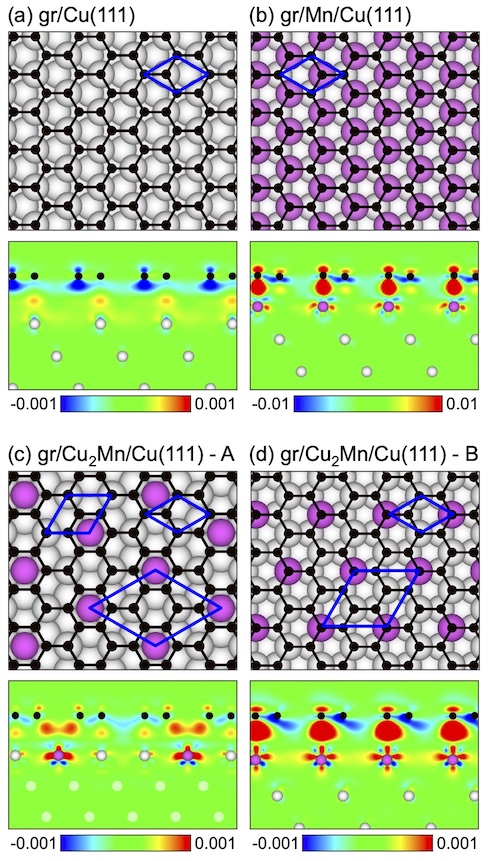}
\caption{Top and side views of different Mn-based intercalation structures: (a) parent graphene/Cu(111), (b) graphene/Mn/Cu(111), (c) graphene/Cu$_2$Mn/Cu(111) -- model A, and (d) graphene/Cu$_2$Mn/Cu(111) --  model B. Spheres of different size and colour represent atoms of different types. Side views for all structures are taken along the graphene arm-chair edge and they are overlaid with electron charge difference maps $\Delta\rho(\mathbf{r}) = \rho_\mathrm{gr/s}(\mathbf{r}) -  \rho_\mathrm{gr}(\mathbf{r}) -  \rho_\mathrm{s}(\mathbf{r})$ (gr: graphene; s: substrate). $\Delta\rho$ is colour coded as blue ($-0.001\,e/\textrm{\AA}^3$), green ($0$), and red ($0.001\,e/\textrm{\AA}^3$) and blue ($-0.01\,e/\textrm{\AA}^3$), green ($0$), and red ($0.01\,e/\textrm{\AA}^3$), for (a,c,d) and (b), respectively. 
}
\label{fig:structure}
\end{figure}

We first discuss the structural and electronic properties of graphene grown on Cu(111) as a reference system. When studying the interface between graphene and close-packed metal surface, three ``high-symmetry'' structures, which preserve $3m$ symmetry, known as \textit{top}--\textit{fcc} (TF), \textit{top}--\textit{hcp} (TH), and \textit{fcc}--\textit{hcp} (FH) are usually considered~\cite{Voloshina:2012bp} (for details, see Supplementary information, Fig.\,S1 and Tab.\,S1). According to our calculations, the TF structure for graphene/Cu(111), schematically shown in Fig.~\ref{fig:structure}(a), was found to be the energetically most stable one. The results of our calculations for this structure are summarized in Tab.~\ref{tab1}. Here, the graphene interaction energy of $-92$\,meV/C-atom and the distance of $3.03$\,\AA\ between graphene and top layer of the Cu(111) slab are indication of the \textit{weak} interaction between graphene and the substrate. Nevertheless, as discussed below, it yields noticeable modifications of the graphene and copper band structures.

\begin{figure}[h]
\centering\includegraphics[width=1.0\textwidth]{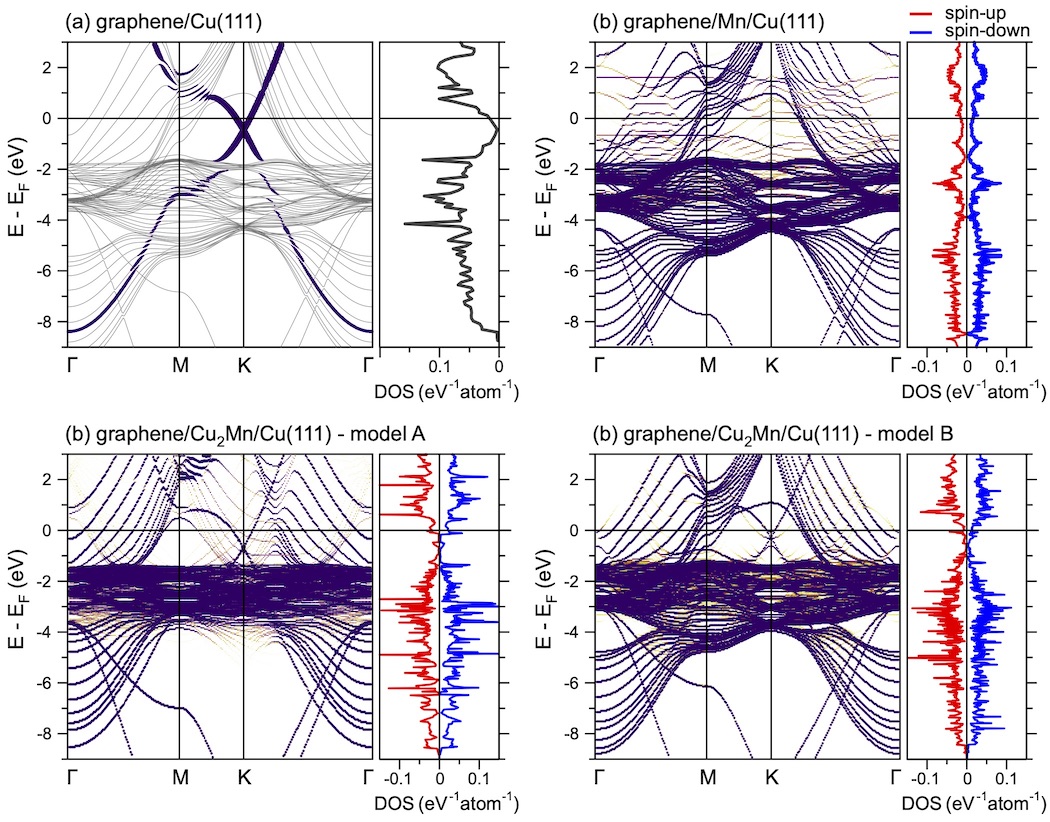}
\caption{Band structures and C-$p_z$ projected density of states calculated for graphene/Cu(111) (a), graphene/Mn/Cu(111) (b),  graphene/Cu$_2$Mn/Cu(111) - model A (c), and graphene/Cu$_2$Mn/Cu(111) - model B (d) in their energetically most favourable structures. In (b-d) spin-resolved band structures obtained after unfolding procedure for the graphene ($1 \times 1$) primitive cell are presented for spin-up channel.}
\label{fig:bands}
\end{figure}

For the clean metal surface, Cu(111), the well-known $L$-gap surface state around the $\Gamma$-point is found in ARPES experiments~\cite{Tamai:2013he}. Using the graphene's lattice constant, the calculated binding energy of this state at the $\Gamma$-point is $-909$\,meV. [Note: A strong deviation from the experimental binding energy is due to the compression of Cu(111) lattice parameters. Complete relaxation of coordinates leads to the in-plane lattice constant of Cu(111) of $2.5241$\,\AA\  and to the significant improvement of agreement between theory ($-471$\,meV) and experiment ($-437$\,meV~\cite{Tamai:2013he})]. In our calculations for graphene/Cu(111) we detect the upward energy shift of the metal surface state by $274$\,meV compared to clean Cu(111). This effect, observed earlier also for the adsorption of graphene on Ag(111) and Au(111) surfaces~\cite{Tesch:2016bd,Tesch:2017jp}, can be explained by the stronger localization of the metal surface state wave-function upon physisorption of graphene on Cu(111) compared to the clean Cu surface. There is also a charge transfer from the Cu(111) surface to graphene (Fig.~\ref{fig:structure}(a)). As a result, graphene became $n$-doped with the position of the Dirac point at $E_D-E_F=-440$\,meV (Fig.~\ref{fig:bands}(a)), which is consistent with the previously published results for graphene/Cu(111)~\cite{Voloshina:2014jl,Gao:2010iz}. Closer look at the calculated band structure of graphene/Cu(111) allows to detect hybridization between the Cu $3d$ and graphene $\pi$ states in the energy range of $E-E_F\approx -2...-4.5$\,eV and consequent opening of the energy gap of $16$\,meV directly at $E_D$~\cite{Voloshina:2014jl}.

When discussing intercalation of Mn under graphene on Cu(111), several scenario are possible. Let us start with an assumption that Mn atoms form a 2D hexagonal layer commensurated with the ($1\times1$)-Cu(111) substrate. Here, Mn intercalant can occupy either \textit{fcc} hollow or \textit{hcp} hollow sites above Cu(111). Together with already mentioned three high-symmetry structures for graphene on Cu(111), this results in $6$ structural possibilities (for details, see Supplementary information, Fig.\,S2 and Tab.\,S2). Besides two possible magnetic orders for Mn atoms -- ferromagnetic (FM) and row-wise antiferromagnetic (AFM) -- have to be taken into account. The energetically most stable structure according to our calculations corresponds to the case where graphene keeps its TF arrangement with respect to Cu(111) and intercalated Mn occupies the \textit{fcc} hollow sites above Cu(111) (Fig.~\ref{fig:structure}(b)). The magnetic moments of Mn atoms ($m=2.07\mu_B$) are coupled antiferromagnetically. The results of our calculations for this structure are summarized in Tab.~\ref{tab1}. Intercalation of Mn yields the increase of the interaction energy between graphene and metal ($E_\mathrm{int}=-237$\,meV/C-atom). This is reflected by the decreased distance between graphene and Mn/Cu(111) ($d_0=2.01$\,\AA), which allows for a strong hybridization between graphene and Mn valence bands states resulting in destruction of the Dirac cone (Fig.~\ref{fig:bands}(b)). The magnetic moment of Mn atoms are $\pm 2.07$ $\mu_{\mathrm{B}}$, while the induced magnetic moments of carbon atoms in graphene overlayer are $\pm 0.01$ $\mu_{\mathrm{B}}$ and $\pm 0.03$ $\mu_{\mathrm{B}}$, which are detectable in magnetic spectroscopic experiments.

According to available experimental and theoretical data, epitaxial Mn films grown on Cu(111) form a surface alloy Cu$_2$Mn having a $(\sqrt{3}\times\sqrt{3})R30^\circ$ superstructure with respect to Cu(111)~\cite{Schneider:1999,Bihlmayer:2000}. Graphene adsorbed on superstructured Mn-Cu(111) surface was previously investigated by means of DFT in the studies on the grain boundaries suppression during the graphene growth~\cite{Chen:2012hw}. In this work the energetically most stable arrangement corresponds to the structure shown in Fig.~\ref{fig:structure}(c) (model A). Here, both monolayers -- the Cu$_2$Mn alloy and the graphene layer -- are rotated by $30^\circ$ with respect to Cu(111). From experimental point of view, the existence of such a structure is hardly possible. Indeed, the preparation of graphene on Mn-Cu(111) is a two-step procedure. Firstly, graphene is synthesized on single crystal Cu(111) by thermal decomposition of, e.\,g., ethylene (C$_2$H$_4$) in an ultrahigh vacuum chamber. As a result, the lattices of graphene and Cu(111) are aligned in the same direction (with a possible formation of the moir\'e structure due to the lattice mismatch between two materials)~\cite{Gao:2010a}. On the second step, the annealing of the prediposited Mn leads to intercalation of Mn with a formation of system, which is supposed to be an ordered graphene/Cu$_2$Mn/Cu(111) trilayer. In such a procedure, it is difficult to expect any change of the graphene orientation (rotation) with respect to the Cu(111) surface. Thus, in our studies, when considering the formation of a Cu$_2$Mn surface alloy at the graphene/Cu(111) interface, the related orientation between graphene and Cu(111) was kept unchanged, while a $(\sqrt{3}\times\sqrt{3})R30^\circ$ superstructure of Cu$_2$Mn with respect to the both Cu(111) and graphene was preserved. For comparison reasons we have considered different high-symmetry arrangements of graphene above the Cu$_2$Mn/Cu(111) substrate and two possible magnetic orders for Mn atoms (for details, see Supplementary information, Tab.\,S3 and Fig.\,S3). The energetically most stable structure -- TH -- is shown in Fig.~\ref{fig:structure}(d) (model B). Here the magnetic moments of Mn atoms are coupled ferromagnetically and the TH and TF arrangements of C-atoms are energetically degenerate with the difference in the total energy of only $2$\,meV (see Tab.\,S3). The calculations for the previously reported structure~\cite{Chen:2012hw} shown in Fig.~\ref{fig:structure}(c) (model A) are also performed for comparison reasons. The results are summarized in Tab.~\ref{tab1}. 

For the both structures under consideration (model A and model B), the calculated interaction energy between graphene and Cu$_2$Mn/Cu(111) as well as the equilibrium distance between graphene and substrate are close to the values obtained for graphene/Cu(111). In both cases graphene is strongly $n$-doped  due to the charge transfer from the metal to graphene valence band states (Fig.~\ref{fig:structure}(c,d)). In both structures, Mn atoms possess quite substantial magnetic moment -- $m_\mathrm{Mn}=3.99\,\mu_\mathrm{B}$ and $m_\mathrm{Mn}=3.58\,\mu_\mathrm{B}$ for model A and model B, respectively (Tab.~\ref{tab1}). However, the induced magnetic moment in a graphene layer does not exceed the value of $0.01\,\mu_\mathrm{B}$ for carbon atoms adsorbed above Mn atoms and they are coupled antiferromagnetically with respect to those of metal ions.

\begin{figure}[h]
\centering\includegraphics[width=1.0\textwidth]{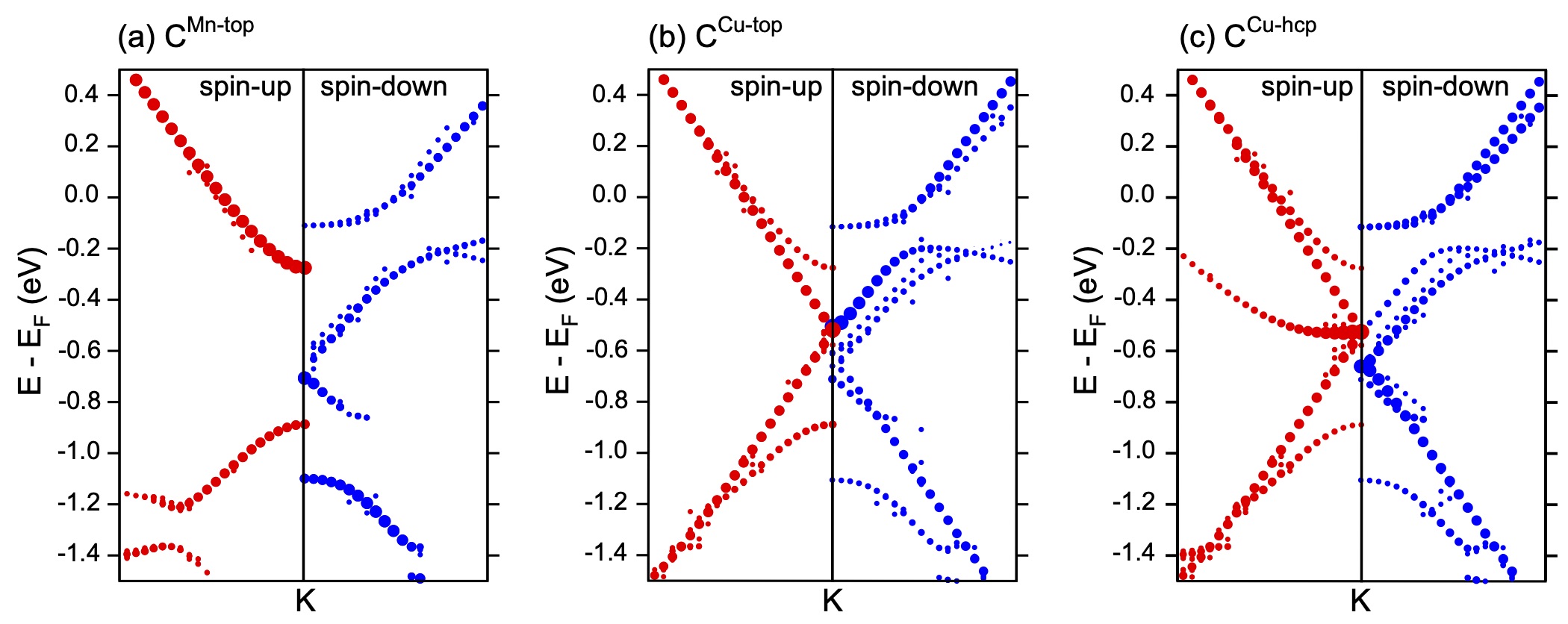}
\caption{Site-decomposition of the valence band states of graphene in the vicinity of the K point as obtained for the graphene/\ch{Cu2Mn}/Cu(111) -- model B structure.}
\label{fig:bands_zoom}
\end{figure}

The calculated band structures (only for spin-up channel) are shown in Fig.~\ref{fig:bands}(c) and Fig.~\ref{fig:bands}(d) for model A and model B, respectively (see also Fig.\,S4 for complete spin-resolved band structures). Fig.~\ref{fig:bands_zoom}(a-c) also shows the zoomed $E-k$ region around the $K$ point for the most energetically stable configuration of the graphene/Cu$_2$Mn/Cu(111) system in model B for spin-up and spin-down channels, respectively. This figure represents the decomposition of the respective spin channels for graphene $\pi$ states around the $K$ point on the C-atoms projected weights corresponding to different carbon atoms in the unit cell of the graphene/Cu$_2$Mn/Cu(111) -- model B structure. From the presented data, it is clear that the existence of three different adsorption sites of carbon atoms in the unit cell for this structure leads to the respective splitting of the graphene $\pi$ band around the $K$ point in to three band's branches. For the first one, associated with the carbon atoms above the Cu-hcp atoms of the Cu(111) slab (Fig.~\ref{fig:bands_zoom}(a)), the linear dispersion of the graphene $\pi$ states is still conserved (although it is spin split) allowing to estimate the doping level of graphene and obtain the position of the Dirac points $E_D-E_F=-0.525$\,eV and $E_D-E_F=-0.660$\,eV for spin-up and spin-down channels, respectively. For carbon atoms adsorbed above the Cu-top atoms of the Cu$_2$Mn layer (Fig.~\ref{fig:bands_zoom}(b)), the linear dispersion is preserved only for spin-up channel and band gap at the Dirac point is opened for the spin-down channel. The most disturbed bands dispersions are found for the graphene bands branches associated with the carbon atoms above the Mn-top atoms of the Cu$_2$Mn layer (Fig.~\ref{fig:bands_zoom}(c)). The significant spin splitting of the bands of $0.215$\,eV in the later case can be assigned to the appearance of the corresponding magnetic moments of carbon atoms adsorbed above the Mn atoms, which have large magnetic moment. The respective energy splitting of the bands and formation of the band gap for these graphene bands branches can be associated with the hybridization of the graphene $\pi$ and Mn\,$3d$ states ($d_{xz}$, $d_{yz}$, $d_{z^2}$). The observed misbalance at the Fermi level of the respective weights for different spin channels of the graphene $\pi$ states might lead to the different conductivity for spin-up and spin-down channels, which could be useful in future spintronics applications and can be a topic for further studies.

\begin{figure}[h]
\centering\includegraphics[width=0.5\textwidth]{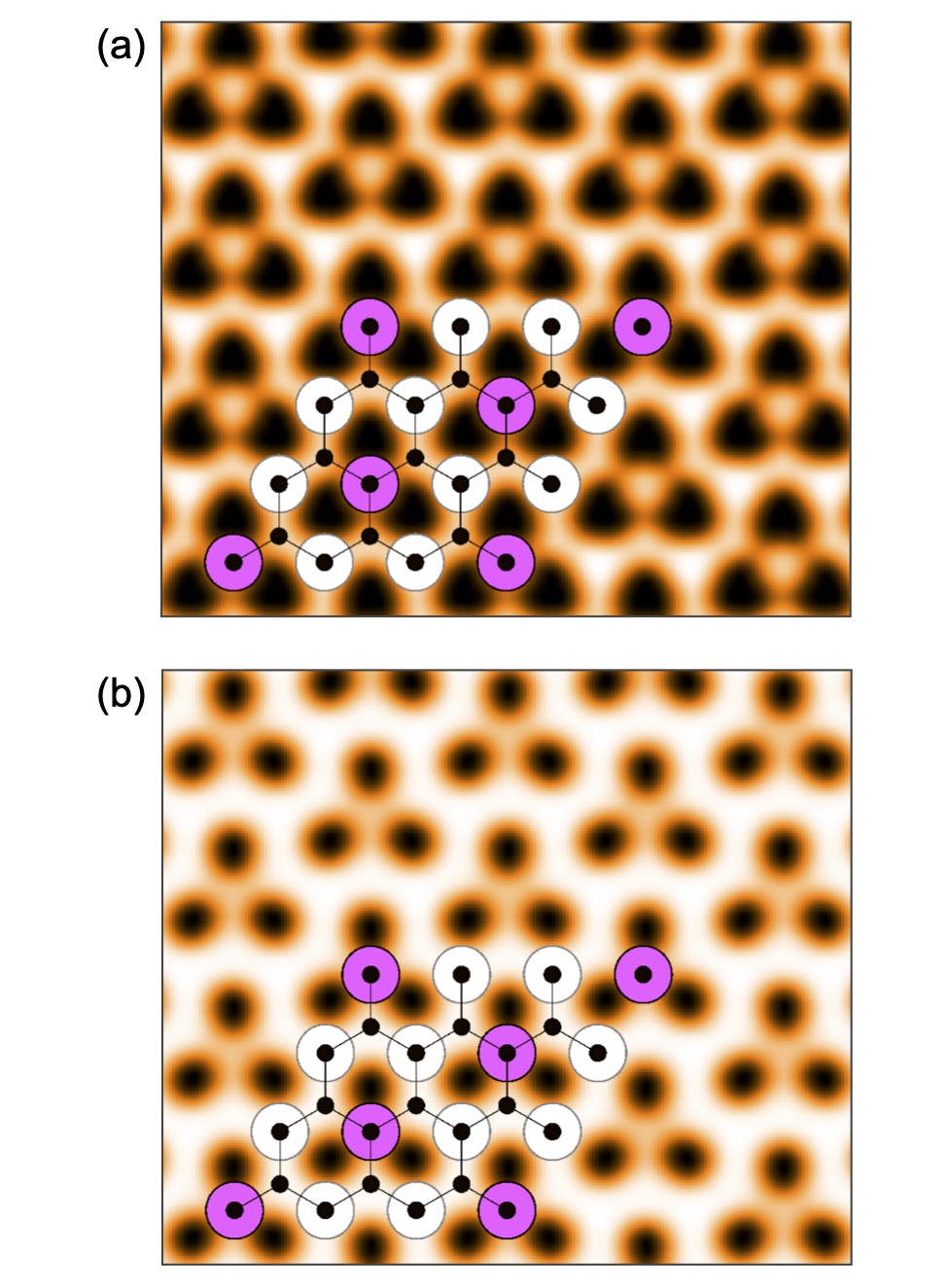}
\caption{Simulated STM images of graphene/\ch{Cu2Mn}/Cu(111) -- model B (a) for occupied ($E-E_F=-100$\,meV) and (b) unoccupied ($E-E_F=+100$\,meV) states, respectively. The images correspond to a tunneling current of $1.2$\,nA~\cite{Hofer:2003}. The STM pictures are overlaid with the crystallographic structure of the studied system.}
\label{fig:STM}
\end{figure}

In Fig.~\ref{fig:STM} we also present calculated STM images of the graphene/Cu$_2$Mn/ Cu(111) -- model B system obtained in the framework of the Tersoff-Hamann approach for electronic states integrated between the Fermi level and energies of (a) $E-E_F=-100$\,meV and (b) $E-E_F=+100$\,meV. Both images demonstrate the appearance of the carbon rings in the STM experiment accompanied by the additional structure due to the symmetry of the Cu$_2$Mn layer at the interface. These data can be used as a good reference in future studies of the structural and electronic properties of the graphene-Mn-Cu(111) system.

\section{Conclusions}
\label{S:Conclu}

Using state-of-the-art DFT calculations the electronic properties of different systems obtained via intercalation of Mn in graphene/Cu(111) were studied. Three models were considered -- sharp graphene/Mn/Cu(111) interface and two configurations for the graphe\-ne/\ch{Cu_2Mn}/Cu(111) Cu-Mn alloy-based systems. In all cases graphene is found as strongly $n$-doped and in case of the sharp graphene/Mn/Cu(111) interface the Dirac cone is fully destroyed. For the most realistic case of the graphene-Mn-Cu(111) system, the formation of the ferromagnetic Cu$_2$Mn alloy at the interface is found. In this case the strong magnetic moment of Mn atoms leads to the appearance of the induced magnetic moment of carbon atoms adsorbed on top of Mn. The formation of the Cu$_2$Mn/Cu(111) slab below graphene also leads to the formation of the interesting band structure and spin topology of the graphene $\pi$ states in the vicinity of the $K$ point. Here, graphene-derived $\pi$ bands are spin split and three branches are formed around the Dirac point, which can be assigned to different carbon sublattices in a graphene layer associated with different carbon atoms adsorption sites above the Cu$_2$Mn/Cu(111) slab. Obtained spin configuration of the graphene $\pi$ bands around the Fermi level could lead to different conductivity for spin-up and spin-down channels, which can be useful for the realization of the spintronics applications on the basis of graphene.

\section{Experimental}
\label{S:Details}

Spin-polarized DFT calculations based on plane-wave basis sets of $500$\,eV cutoff energy were performed with the Vienna \textit{ab initio} simulation package (VASP)~\cite{Kresse:1996kg,Kresse:1994cp,Kresse:1993hw}. The Perdew-Burke-Ernzerhof (PBE) exchange-correlation functional~\cite{Perdew:1997ky} was employed. The electron-ion interaction was described within the projector augmented wave (PAW) method~\cite{Blochl:1994fq} with C ($2s$,$2p$), Mn ($3p$,$3d$,$4s$), and Cu ($3d$,$4s$) states treated as valence states. The Brillouin-zone integration was performed on $\Gamma$-centered symmetry reduced Monkhorst-Pack meshes using a Methfessel-Paxton smearing method of first order with $\sigma=0.15$\,eV, except for the calculation of total energies. For these calculations, the tetrahedron method with Bl\"ochl corrections~\cite{Blochl:1994ip} was employed. The $k$ mesh for sampling the supercell Brillouin zone are chosen to be as dense as at least $24\times 24$, when folded up to the simple graphene unit cell. Dispersion interactions were considered adding a $1/r^6$ atom-atom term as parameterized by Grimme (``D2'' parameterization)~\cite{Grimme:2006fc}. During structure optimization, the convergence criteria for energy was set equal to $10^{-6}$\,eV.  The band structures calculated for the studied systems were unfolded (if necessary) to the graphene ($1\times 1$) primitive unit cell according to the procedure described in Refs.~\citenum{Medeiros:2014ka} and \citenum{Medeiros:2015ht} with the code BandUP. The STM images are calculated using the Tersoff-Hamann formalism~\cite{Tersoff:1985}.

The studied interfaces are modeled by a slab consisting of $10$ Cu-layers, a graphene adsorbed on the top side of the slab, and a vacuum gap of at least $20$\,\AA. The bottom layer of the slab is protected by H-atoms. In the case of the graphene-Mn-Cu(111) system, a Mn layer is added between graphene and Cu(111). In order to describe graphene as realistically as possible, the lattice constant of copper is set to be compatible with the optimized graphene lattice constant ($a = 2.4636$\,\AA). During the structural optimization, all the $z$-coordinates of carbon atoms ($x$ and $y$-coordinates are fixed) as well as those of the top three layers of metal atoms ($z$-coordinates) are relaxed until forces became smaller than $2\times10^{-2}$\,eV/\AA.

\begin{acknowledgement}

This work was supported by the National Natural Science Foundation of China (Grant No. 21973059). 

\end{acknowledgement}

\begin{suppinfo}


The following files are available free of charge:

Additional theoretical data for the graphene-Cu(111) and graphene-Mn-Cu(111) systems.


\end{suppinfo}


\clearpage
\begin{table}
  \caption{Results for the atomic structure of the graphene/metal interface models and for the clean metal surfaces: $E_\mathrm{int}$ (in meV/C-atom) is the interaction energy per carbon atom (There are two carbon atoms in graphene ($1\times1$) unit cell); $d_0$ (in \AA) is the mean distance between the graphene overlayer and the interface metal layer; $d_1$ (in \AA) is the mean distance between the interface metal layer and the second metal layer; $d_2$ (in \AA) is the mean distance between the second and third metal layers; $m_\mathrm{Mn}$  ($\mu_\mathrm{B}$) is the interface Mn spin magnetic moment; $m_\mathrm{C}$ (in $\mu_\mathrm{B}$) is the interface carbon spin magnetic moment (several values for the nonequivalent carbon atoms are indicated).}
  \label{tab1}
\begin{tabular}{l c c c c}
    \hline
System 	& graphene/Cu(111) 		& graphene/Mn/Cu(111)	& \multicolumn{2}{c}{graphene/Cu$_2$Mn/Cu(111)} \\
		&					&					& model A		&model B     \\ 
    \hline
$E_\mathrm{int}$&$-92$  					&$-237$      			& $-84$  			& $-89$                     \\
$d_0$			&$3.03$				&$2.01$  				& 2.99 			&  3.05						\\
$d_1$			&$2.10$				&$2.14$  				& 1.92 			&  2.16						\\
$d_2$			&$2.09$				&$2.10$ 	 			& 1.88 			&  2.09						\\
$m_\mathrm{Mn}$	&---					&$\pm 2.07$  			& 3.99 			&  3.58					\\
$m_\mathrm{C}$	&---					&$\pm 0.01/\pm 0.03$	& 0.00 			& $0.00$/$0.00$/$-0.01$				\\
$E_D-E_F$		&$-440$                 		&   ---                    		& $-650$                	& $-525$/$-660$                    \\
    \hline
  \end{tabular}
\end{table}

\clearpage

\clearpage

\clearpage

\clearpage

\clearpage

\noindent
\textbf{Supplementary materials for ``Intercalation of Mn in graphene/Cu(111) interface: Insights to the electronic and magnetic properties from theory''}






\subsection*{List of Tables and Figures:}

\medskip
\noindent \textbf{Tab.\,S1:} Results for the relaxed graphene/Cu(111). $\Delta E$ (in meV/u.c.) is the relative total energies per graphene $(1\times1)$ unit cell (u.c.) with respect to the energetically most favourable structure. $d_0$ (in \AA) is the distance between a graphene overlayer and the interface metal layer; $d_1$ (in \AA) is the mean distance between the interface metal layer and the second metal layer; $d_2$ (in \AA) is the mean distance between the second and third metal layers.

\medskip
\noindent \textbf{Tab.\,S2:} Results for the relaxed graphene/Mn/Cu(111) with different magnetic configuration (FM=ferro\-magnetic, AFM=anti-ferro\-magnetic). $\Delta E$ (in meV/u.c.) is the relative total energies per graphene $(1\times1)$ unit cell (u.c.) with respect to the energetically most favourable structure. $d_0$ (in \AA) is the distance between the graphene overlayer and the interface metal layer; $d_1$ (in \AA) is the mean distance between the interface metal layer and the second metal layer; $d_2$ (in \AA) is the mean distance between the second and third metal layers; $m_{\mathrm{C}}$ and $m_{\mathrm{Mn}}$ (in $\mu_{\mathrm{B}}$) is the interface/surface carbon atoms and manganese atoms spin magnetic moments.

\medskip
\noindent \textbf{Tab.\,S3:} Results for the relaxed graphene/Cu$_2$Mn/Cu(111) (model B) with different magnetic configuration (FM and AFM). Here $\Delta E$ (in meV/s.c.) is the relative total energies per graphene $(3\times3)$ super cell (s.c.) with respect to the energetically most favourable structure.

\medskip
\noindent \textbf{Fig.\,S1:} Top view of the crystallographic structures of graphene/Cu(111): (a) TF - the C atoms are placed directly above the Cu atoms of the first layer (\emph{top} site) and the third layer (\emph{fcc} site); (b) TH - the C atoms are placed directly above the Cu atoms of the first layer (\emph{top} site) and the second layer (\emph{hcp} site); (c) FH - the C atoms are placed directly above the Cu atoms of the third layer (\emph{fcc} site) and the second layer (\emph{hcp} site). The graphene units cell is marked with the blue rhombus.

\medskip
\noindent \textbf{Fig.\,S2:} Top view of the crystallographic structures of graphene/Mn/Cu(111) where the Mn atoms can be located at the \emph{fcc} or \emph{hcp} site of Cu(111) surface, respectively while the two inequivalent carbon atoms of graphene can adopt the FH, TH or TF configuration. The graphene units cell is marked with the blue rhombus.

\medskip
\noindent \textbf{Fig.\,S3:} Surface alloy model of graphene/Cu$_2$Mn/Cu(111) system (model B) with \emph{hcp} lattice, where the first layer of Cu(111) slab is replaced by \ch{Cu2Mn} surface alloy and the two inequivalent carbon atoms of graphene can adopt the FH, TH or TF configuration.

\medskip
\noindent \textbf{Fig.\,S4:} Spin-resolved band structures obtained after unfolding procedure for the graphene ($1\times 1$) primitive cell for (a) graphene/Cu$_2$Mn/Cu(111) - model A and (b) graphene/Cu$_2$Mn/Cu(111) - model B in their energetically most favourable structures.

\newpage

\begin{table}[!htp]
\caption{Results for the relaxed graphene/Cu(111). $\Delta E$ (in meV/u.c.) is the relative total energies per graphene $(1\times1)$ unit cell (u.c.) with respect to the energetically most favourable structure, $E_{\mathrm{int}}$ is the interaction energy of graphene and substrate. $d_0$ (in \AA) is the distance between a graphene overlayer and the interface metal layer; $d_1$ (in \AA) is the mean distance between the interface metal layer and the second metal layer; $d_2$ (in \AA) is the mean distance between the second and third metal layers.}
\centering
\medskip
\small
\begin{tabular}{c|c|c|c|c|c|l}
\hline
Structure 	& $\Delta E$ 	& $E_{\mathrm{int}}$ & $d_0$ 		& $d_1$ 	& $d_2$ &	 Reference \\[0.2cm]  \hline
TF		&$0.00$		&$-92$			&$3.03$		&$2.10$	&$2.09$	& Fig.~\ref{fig:geo_grCu111}a\\[0.2cm]
TH		&$2.26$		&$-92$			&$3.03$		&$2.11$	&$2.09$	& Fig.~\ref{fig:geo_grCu111}b\\[0.2cm]
FH		&$11.89$		&$-87$			&$3.07$		&$2.11$	&$2.09$	& Fig.~\ref{fig:geo_grCu111}c\\[0.2cm]
\hline
\end{tabular}
\end{table}

\newpage

\begin{table}[!htp]
\caption{Results for the relaxed graphene/Mn/Cu(111) with different magnetic configuration (FM=ferro\-magnetic, AFM=anti-ferro\-magnetic). $\Delta E$ (in meV/u.c.) is the relative total energies per graphene $(1\times1)$ unit cell (u.c.) with respect to the energetically most favourable structure, $E_{\mathrm{int}}$ is the interaction energy of graphene and substrate. $d_0$ (in \AA) is the distance between the graphene overlayer and the interface metal layer; $d_1$ (in \AA) is the mean distance between the interface metal layer and the second metal layer; $d_2$ (in \AA) is the mean distance between the second and third metal layers; $m_{\mathrm{C}}$ and $m_{\mathrm{Mn}}$ (in $\mu_{\mathrm{B}}$) is the interface/surface carbon atoms and manganese atoms spin magnetic moments.}
\centering
\medskip
\small
\begin{tabular}{c|c|c|c|c|c|c|c|c|c|l}
\hline
Magnetic          		& \multicolumn{2}{c|}{Structures} & $\Delta E$ & $E_{\mathrm{int}}$ & $d_0$ & $d_1$ & $d_2$ & $m_{\mathrm{Mn}}$ & $m_\mathrm{C}$ &Reference \\\cline{2-3}
  state     			& gr & Mn 		&  		&  		& 		&  			&  		&&  \\ \hline
  \multirow{6}*{FM}	& \multirow{2}*{FH} 	& hcp 	&$110$	& ---  	& 2.02/1.94 	& 2.14 	& 2.10 & 1.62 & $-0.02/-0.01$&Fig.~\ref{fig:geo_grMnCu111}d \\
   				&                   		& fcc 	&$98$	& ---  	& 2.02/1.94 	& 2.14 	& 2.10 & 1.63 & $-0.01/-0.02$&Fig.~\ref{fig:geo_grMnCu111}a \\ \cline{2-10}
   				& \multirow{2}*{TF} 	& hcp 	&$717$	& ---  	& 1.93/1.93 	& 2.13 	& 2.10 & 1.68 & $-0.02/-0.02$&Fig.~\ref{fig:geo_grMnCu111}e \\
   				&                   		& fcc 	&$58$	&$-360$	& 2.02/1.94 	& 2.13 	& 2.10 & 1.62 & $-0.01/-0.02$&Fig.~\ref{fig:geo_grMnCu111}b \\ \cline{2-10}
   				& \multirow{2}*{TH} 	& hcp 	&$67$ 	& ---  	& 2.02/1.94 	& 2.13 	& 2.11 & 1.61 & $-0.01/-0.02$&Fig.~\ref{fig:geo_grMnCu111}f \\
   				&                   		& fcc 	&$713$	& ---  	& 1.93/1.93 	& 2.12 	& 2.10 & 1.70 & $-0.02/-0.02$&Fig.~\ref{fig:geo_grMnCu111}c \\ \hline
  \multirow{6}*{AFM}& \multirow{2}*{FH} 	& hcp 	&$217$ 	& ---  	& 2.04/1.98 	& 2.15 	& 2.10 & $\pm 0.04$ & 0.00/0.00&Fig.~\ref{fig:geo_grMnCu111}d \\
   				&                   		& fcc 	&$194$ 	& ---  	& 2.01/1.93 	& 2.10 	& 2.09 & $\pm 0.02$ & 0.00/0.00&Fig.~\ref{fig:geo_grMnCu111}a \\ \cline{2-10}
   				& \multirow{2}*{TF} 	& hcp 	&$302$ 	& ---  	& 3.22/3.22 	& 2.18 	& 2.11 & $\pm 2.89$ & 0.00/0.00&Fig.~\ref{fig:geo_grMnCu111}e \\
   				&                   		& fcc 	&$0$ 	&$-237$	& 2.04/1.98 	& 2.14 	& 2.10 & $\pm 2.07$ & $\pm 0.01/\pm 0.03$& Fig.~\ref{fig:geo_grMnCu111}b \\ \cline{2-10}
   				& \multirow{2}*{TH} 	& hcp 	&$212$ 	& ---  	& 2.04/1.99 	& 2.15 	& 2.11 & $\pm 0.07$ & 0.00/0.00&Fig.~\ref{fig:geo_grMnCu111}f \\
   				&                   		& fcc 	&$294$ 	& ---  	& 3.29/3.29 	& 2.19 	& 2.11 & $\pm 2.90$ & 0.00/0.00&Fig.~\ref{fig:geo_grMnCu111}c \\
\hline
\end{tabular}
\label{tbl:grMnCu}
\end{table}

\newpage

\begin{table}[!htp]
\caption{Results for the relaxed graphene/Cu$_2$Mn/Cu(111) (model B) with different magnetic configuration (FM and AFM). Here $\Delta E$ (in meV/s.c.) is the relative total energies per graphene $(3\times3)$ super cell (s.c.) with respect to the energetically most favourable structure, $E_{\mathrm{int}}$ is the interaction energy of graphene and substrate.}
\centering
\medskip
\small
\begin{tabular}{c|c|c|c|c|c|c|c|c| l}
\hline
                 	 	&Magnetic& $\Delta E$ 	& $E_{\mathrm{int}}$ & $d_0$ & $d_1$ & $d_2$ & $m_{\mathrm{Mn}}$ & $m_{\mathrm{C}}$ &Reference\\ 
                  		&state&				&			&		&		&	&		& &\\ \hline
\multirow{2}*{FH} 	& AFM 	&$106$ 		& ---      		& $3.15$ 	& $2.18$ 	& $2.10$ & $\pm3.69/\pm3.72/\pm3.70$ & 0.00 &\multirow{2}*{Fig.~\ref{fig:model_B}a}\\
                  		&  FM 	&$54$ 		& ---      		& $3.11$ 	& $2.17$ 	& $2.09$ & $3.63		 $ &  0.00 & \\ \hline
\multirow{2}*{TH} 	& AFM 	&$34$		& ---     	 	& $3.10$ 	& $2.17$ 	& $2.10$ & $\pm3.67/\pm3.69/\pm3.67$ & 0.00 &\multirow{2}*{Fig.~\ref{fig:model_B}b} \\
                  		&  FM 	&$0$ 		& $-89.23$	& $3.05$ 	& $2.16$ 	& $2.09$ & $3.58		 $ &0.00 & \\ \hline
\multirow{2}*{TF} 	& AFM 	&$28$ 		& $-87.28$ 	& $3.10$ 	& $2.17$ 	& $2.10$ & $\pm3.66/\pm3.69/\pm3.67$ &0.00 &\multirow{2}*{Fig.~\ref{fig:model_B}c}  \\
                  		&  FM 	&$2$			& $-88.95$ 	& $3.05$ 	& $2.16$ 	& $2.09$ & $3.58		 $ &  0.00 &\\
\hline
\end{tabular}
\label{tbl:B}
\end{table}

\newpage

\begin{figure}[!htp]
\centering
\includegraphics[width=0.8\textwidth]{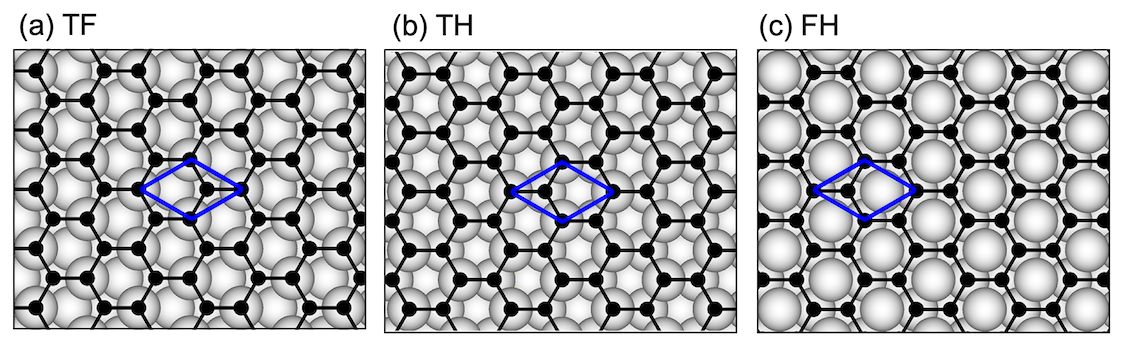}

\label{fig:geo_grCu111}
\end{figure}
\noindent Fig.\,S1. Top view of the crystallographic structures of graphene/Cu(111): (a) TF - the C atoms are placed directly above the Cu atoms of the first layer (\emph{top} site) and the third layer (\emph{fcc} site); (b) TH - the C atoms are placed directly above the Cu atoms of the first layer (\emph{top} site) and the second layer (\emph{hcp} site); (c) FH - the C atoms are placed directly above the Cu atoms of the third layer (\emph{fcc} site) and the second layer (\emph{hcp} site). The graphene units cell is marked with the blue rhombus.

\newpage

\begin{figure}[!htp]
\centering
\includegraphics[width=0.8\textwidth]{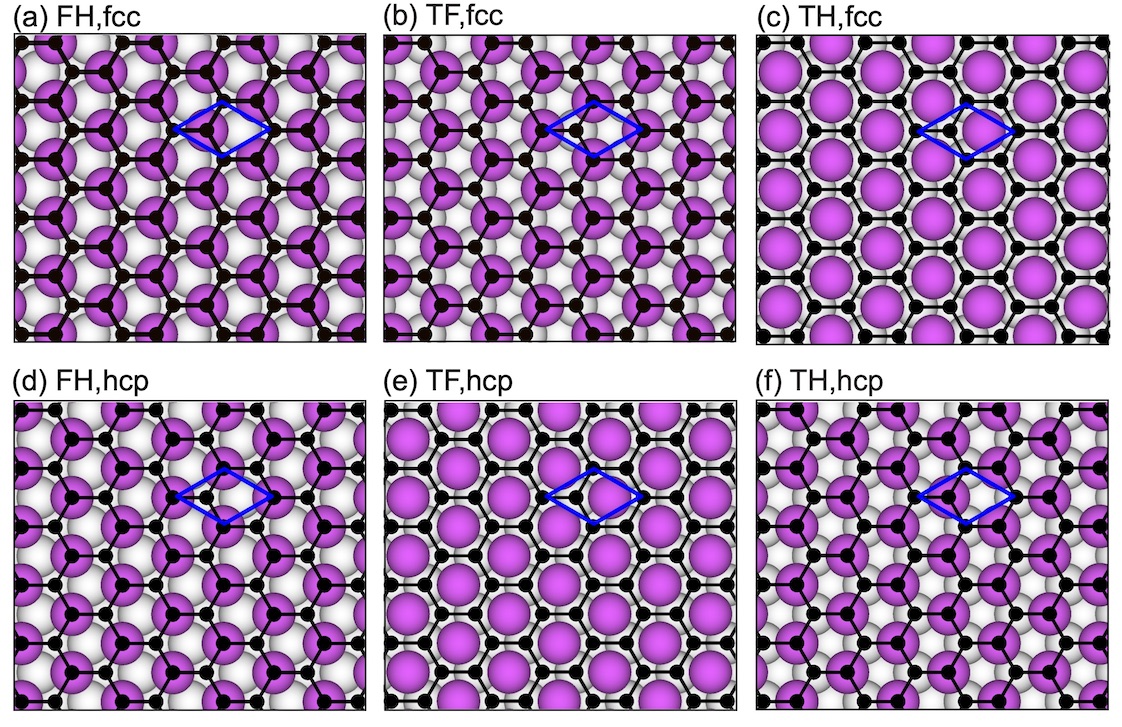}

\label{fig:geo_grMnCu111}
\end{figure}
\noindent Fig.\,S2. Top view of the crystallographic structures of graphene/Mn/Cu(111) where the Mn atoms can be located at the \emph{fcc} or \emph{hcp} site of Cu(111) surface, respectively while the two inequivalent carbon atoms of graphene can adopt the FH, TH or TF configuration. The graphene units cell is marked with the blue rhombus.

\newpage

\begin{figure}[!htp]
\centering
\includegraphics[width=0.8\textwidth]{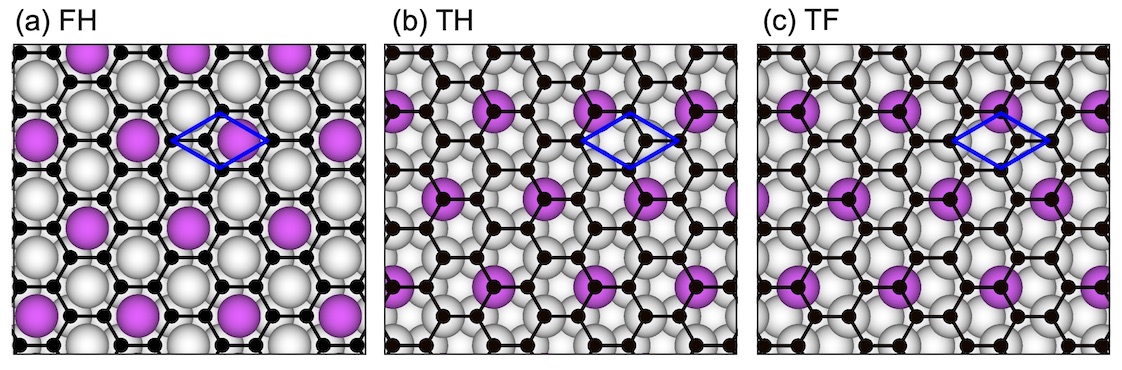}

\label{fig:model_B}
\end{figure}
\noindent Fig.\,S3. Surface alloy model of graphene/Cu$_2$Mn/Cu(111) system (model B) with \emph{hcp} lattice, where the first layer of Cu(111) slab is replaced by Cu$_2$Mn surface alloy and the two inequivalent carbon atoms of graphene can adopt the FH, TH or TF configuration.

\newpage

\begin{figure}
\centering
\includegraphics[width=0.8\textwidth]{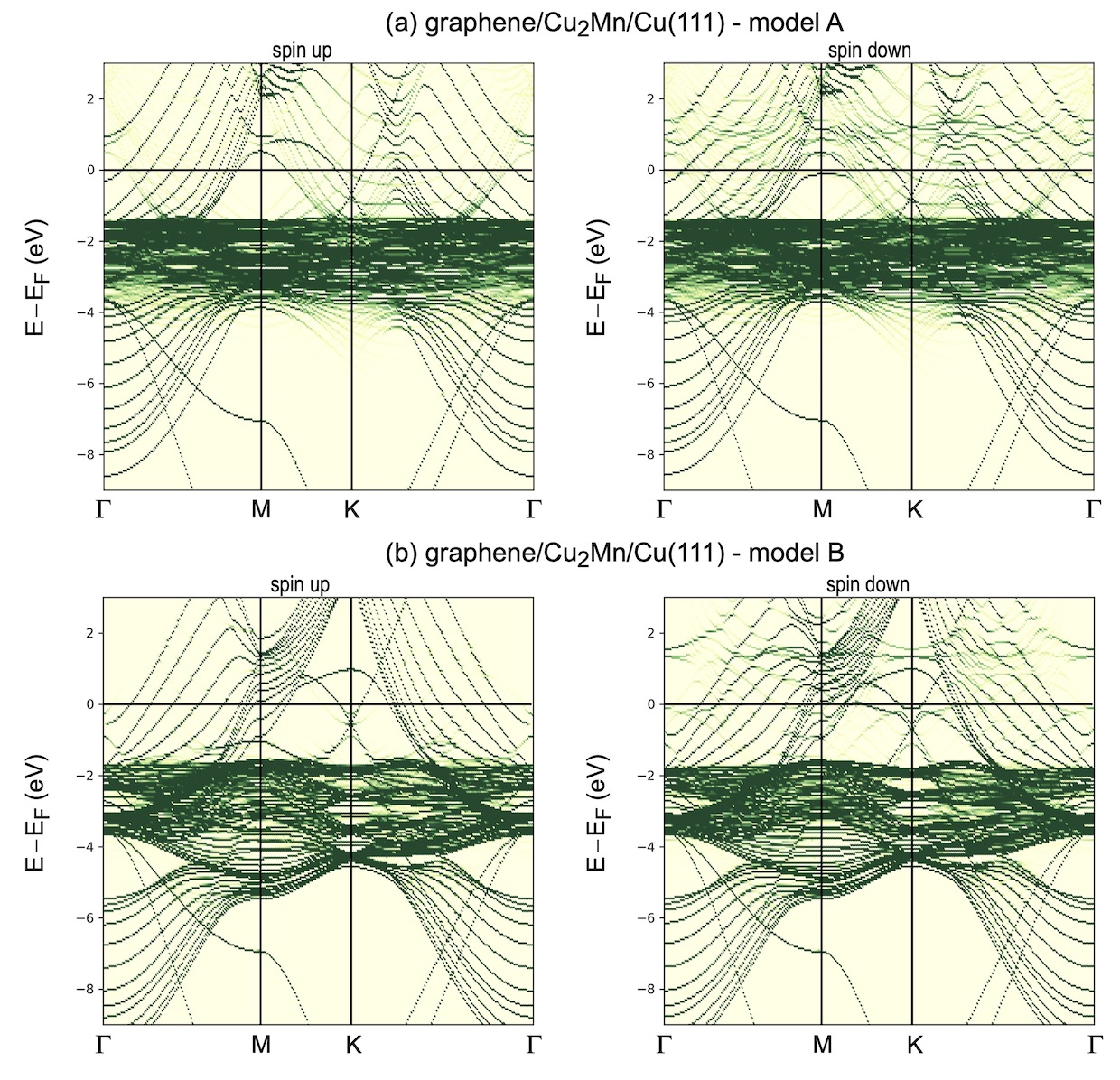}

\label{figs:bands}
\end{figure}
\noindent Fig.\,S4. Spin-resolved band structures obtained after unfolding procedure for the graphene ($1\times 1$) primitive cell for (a) graphene/Cu$_2$Mn/Cu(111) - model A and (b) graphene/Cu$_2$Mn/Cu(111) - model B in their energetically most favourable structures.

\end{document}